\documentclass[aps,prc,preprint,groupedaddress,floatfix,showpacs,twoside]{revtex4}
\usepackage{graphicx}
\usepackage{dcolumn}
\usepackage{bm}
\begin{document}

\title{Liquid-Gas Coexistence and Critical Behavior in
Boxed Pseudo-Fermi Matter}

\author{Jan T\~oke}
\author{Jun Lu}
\author{W.Udo Schr\"oder}
\affiliation{Department of Chemistry, University of Rochester,
Rochester, New York 14627}

\date{\today}

\begin{abstract}
A schematic model is presented that allows one to study the
behavior of interacting pseudo-Fermi matter, locked in a
thermostatic box. As a function of the box volume and temperature,
the matter is seen to show all of the familiar characteristics of
a Van der Waals gas, which include the coexistence of two phases
under certain circumstances and the presence of a critical point.

\end{abstract}


\maketitle

\section{Introduction}

The possibility that a liquid-gas phase transition in finite
nuclei may manifest itself via copious production of
intermediate-mass fragments (IMF) in energetic heavy-ion reactions
has driven both, theoretical and experimental studies of nuclear
multifragmentation over more than a decade. A prominent role in
theoretical considerations is played by the concept of a freezeout
configuration.\cite{mmmc,smm} This concept implies the existence
of a definite volume, within which the system reaches a state
close to thermal equilibrium. While the existence of an effective
freezout volume may be debatable, the concept itself is useful for
understanding the possible behavior of nuclear matter under
various conditions. With such a didactic strategy in mind, and as
an extension of earlier studies modelling the behavior of finite
Fermi systems, \cite{toke_quantum,toke_surface,toke_surfentr} the
present study considers pseudo-Fermi matter confined to a box of
definite volume. It evaluates the isothermal behavior of such
matter and its dependence on box volume and temperature. The
utmost simplicity of the formalism allows one to gain insight into
physical phenomena that may be obscured in more rigorous
approaches.

\section{Theoretical Formalism}
\label{sec:formalism}

The present study considers a scenario of nuclear pseudo-Fermi
matter of mass number $A$, locked in a spherical box of a volume
$V$ and kept at constant temperature $T$. In the proposed
formalism, thermostatic properties of the matter are modelled by
two equations, the isochoric caloric equation of state, and the
zero-temperature equation of state. The equilibrium state of the
system is then found as a function of $V$ and $T$, based on the
requirement that the free energy of the system be minimal.

First, we consider homogeneous systems with constant density
throughout their volume, i.e., systems in states where only one
phase is present. The isochoric caloric equation of state for such
a system is taken in a simple form adequate for low-temperature
Fermi gases

\begin{equation}
\label{eq:caloric_eq} E^*_{therm}=aT^2,
\end{equation}

\noindent where $E^*_{therm}$ is the thermal energy and $a$ is the
level density parameter. The latter parameter is assumed to depend
on matter density as

\begin{equation}
\label{eq:little_a} a=a_o({\rho\over \rho_o})^{-{2\over 3}},
\end{equation}
\noindent in accordance with the low-temperature Fermi gas model.

It is worthwhile keeping in mind, that for Fermi gases,
Eq.~\ref{eq:caloric_eq} is a good approximation only for
temperatures that are small compared to the Fermi energy, i.e.,
for $T<<E_F$. Notably, for diluted or very hot Fermi systems, with
the matter density approaching zero, $\rho->0$, or for high
temperatures, $T->\infty$, the caloric equation of state
approaches asymptotically that of a classical gas:

\begin{equation}
\label{eq:caloric_eq_classical} E^*_{therm}={3\over 2}AT
\end{equation}

For the sake of simplicity and without loss of generality, the
present study uses Eq.~\ref{eq:caloric_eq} over the full range of
matter densities and temperatures considered. The term
``pseudo-Fermi'' matter is used to distinguish the matter
considered here from true Fermi matter.

The second defining equation, the zero-temperature equation of
state, expresses the compressional (potential) energy of the
system as a function of the system volume or as a function of
matter density. The present study adopts the harmonic
approximation in the form used in the Expanding Emitting Source
Model (EESM) \cite{eesm}, whereby the in-medium nucleonic
(potential) energy changes quadratically with the relative
deviation of the actual matter density from the ground-state
density, by an amount

\begin{equation}
\label{eq:compress} \epsilon_{compr}=-\epsilon_B(1-{\rho\over
\rho_o})^2.
\end{equation}

In Eq.~\ref{eq:compress}, $\epsilon_{compr}$ and $\epsilon_B$ are
compressional and ground-state binding energies per nucleon,
respectively, and $\rho$ and $\rho_o$ are the actual and the
ground-state matter densities, respectively. Equation
\ref{eq:compress} implies an effective ground-state
incompressibility constant of $K_o$=-18$\epsilon_B$. Assuming
$\epsilon_B$=-8 MeV, the effective incompressibility, including
the effects of surface tension, equals $K_o=144$ MeV. Note, that
within the harmonic approximation, for infinite nuclear matter
characterized by $\epsilon_B\approx -16$ MeV, the
incompressibility constant is $K_o \approx 288$ MeV. The latter
value places the present harmonic approximation ``neutrally''
between the currently considered limits of ``soft'' and ``hard''
equations of state for nuclear matter.

While the two defining equations \ref{eq:caloric_eq} and
\ref{eq:compress} may be considered rather crude approximations,
they do contain the essential physics responsible for first-order
phase transitions and critical phenomena. Given these two
equation, one can write expressions for all thermodynamic
quantities characterizing the system, including the Helmholtz free
energy $F$. The state of the system can then be found by
minimizing the free energy, for any box volume $V$ and temperature
$T$.

Based on Eq.~\ref{eq:caloric_eq}, one can write for the entropy
$S$ of a homogeneous system
\begin{equation}
\label{eq:entropy} S=\int_0^{E*}{1\over T}d\epsilon=2\sqrt{aE^*}.
\end{equation}

The free energy, $F$, for a homogeneous, single-phase system is
given by
\begin{equation}
\label{eq:free_energy}
F=E^*_{total}-ST=E^*_{compr}+E^*_{therm}-2aT^2=E^*_{compr}-aT^2.
\end{equation}

In Eq.~\ref{eq:free_energy}, for sake of simplicity, the free
energy is given relative to a constant ground state energy of the
system. Further, using Eqs.~\ref{eq:little_a} and
\ref{eq:compress}, Eq.~\ref{eq:free_energy} can be rewritten in a
form revealing explicitly the important dependence on the matter
density $\rho$, i.e.,
\begin{equation}
\label{eq:free_energy_f} F=E_B(1-{\rho\over
\rho_o})^2-a_o({\rho\over \rho_o})^{-{2\over 3}}T^2.
\end{equation}

Eq.~\ref{eq:free_energy_f} allows one to write equations for the
system pressure $p$ and the chemical potential $\mu$ as functions
of volume (matter density) and temperature.

The pressure $p$ can be expressed generally as the negative
partial derivative of the free energy $F$ with respect to volume
$V$, at fixed number of nucleons $A$ and fixed temperature $T$,
i.e.,
\begin{equation}
\label{eq:pressure_general} p=-({\partial F\over \partial
V})_{A,T}={1\over A}\rho^2({\partial F\over\partial \rho})_{A,T}.
\end{equation}

Thus, for the case of pseudo-Fermi matter considered here, one
obtains based on Eqs.~\ref{eq:free_energy_f} and
\ref{eq:pressure_general}
\begin{equation}
p=2\epsilon_B\rho_o(1-{\rho\over \rho_o})({\rho\over
\rho_o})^2+{2\over 3}\alpha_o\rho_o({\rho\over \rho_o})^{1\over
3}T^2, \label{eq:pressure_pseudo}
\end{equation}

\noindent where $\epsilon_B$ and $\alpha_o$ are the binding energy
per nucleon and the level-density parameter per nucleon,
respectively.

The chemical potential $\mu$ can be expressed generally as a
partial derivative of free energy $F$ with respect to the number
of nucleons $A$, taken at fixed volume $V$ and fixed temperature
$T$, i.e.,
\begin{equation}
\label{eq:chempot_general} \mu=({\partial F\over \partial
A})_{V,T}.
\end{equation}

Using Eqs.~\ref{eq:free_energy_f} and \ref{eq:chempot_general},
and noting further that $\rho=A/V$, $E_B=A\epsilon_B$ and
$a_o=A\alpha_o$, one obtains for the chemical potential
\begin{equation}
\label{eq:chempot_pseudo} \mu=\epsilon_B[1-4{\rho\over
\rho_o}+3({\rho\over \rho_o})^2]-{1\over 3}\alpha_o({\rho\over
\rho_o})^{-{2\over 3}}T^2.
\end{equation}

It is worth noting in Eq.~\ref{eq:chempot_pseudo} that, as a
result of the requirement for $V$ to be constant, the magnitude of
the chemical potential differs significantly from the value of the
free energy per nucleon. For example, the contribution of thermal
excitation to the chemical potential is only one-third of what
constitutes the thermal part of the free energy per nucleon.

\section{Liquid-Gas Coexistence}

When confined to a thermostatic box, matter will eventually fill
all of the available volume $V$ so as to minimize the free energy
of the system. For non-interacting matter, the minimum free energy
always corresponds to a uniform density distribution. This is
generally not true for an interacting system. In particular, for
interacting pseudo-Fermi matter at the minimum free energy,
low-density and high-density phases coexist.

For a two-phase system of $A$ nucleons confined to volume $V$ at
temperature $T$, the free energy can be expressed as a function of
two variables, e.g., in terms of the volume of the gaseous phase,
$V_{gas}$, and the number of nucleons contained in this phase,
$A_{gas}$.
\begin{equation}
\label{eq:freen_2phase} F=F_{gas}+F_{liquid},
\end{equation}

Inserting for $F_{gas}$ and $F_{liquid}$ the expressions given by
Eq.~\ref{eq:free_energy_f} for the corresponding numbers of
nucleons $A_{gas}$ and $A_{liquid}$, respectively, one obtains

\begin{equation}
\label{eq:free_energy_gas}
F_{gas}=A_{gas}\epsilon_B(1-{A_{gas}\over
V_{gas}\rho_o})^2-A_{gas}\alpha_o({A_{gas}\over
V_{gas}\rho_o})^{-{2\over 3}}T^2
\end{equation}

\noindent and
\begin{equation}
\label{eq:free_energy_liquid}
F_{liquid}=(A-A_{gas})\epsilon_B[1-{A-A_{gas}\over
(V-V_{gas})\rho_o}]^2-(A-A_{gas})\alpha_o[{A-A_{gas}\over
(V-V_{gas})\rho_o}]^{-{2\over 3}}T^2.
\end{equation}

The condition for the minimum free energy can be expressed in form
of two equations, reflecting requirements of dynamical and
chemical equilibrium, respectively
\begin{equation}
\label{eq:dyn_equil} ({\partial F\over \partial
V_{gas}})_{T,A_{gas}}=({\partial F_{gas}\over \partial
V_{gas}})_{T,A_{gas}}-({\partial F_{liquid}\over \partial
V_{liquid}})_{T,A_{liquid}}=p_{gas}-p_{liquid}=0
\end{equation}
\noindent and
\begin{equation}
\label{eq:chem_equil} ({\partial F\over \partial
A_{gas}})_{T,V_{gas}}=({\partial F_{gas}\over \partial
A_{gas}})_{T,V_{gas}}-({\partial F_{liquid}\over \partial
A_{liquid}})_{T,V_{liquid}}=\mu_{gas}-\mu_{liquid}=0.
\end{equation}

\begin{figure}
\includegraphics{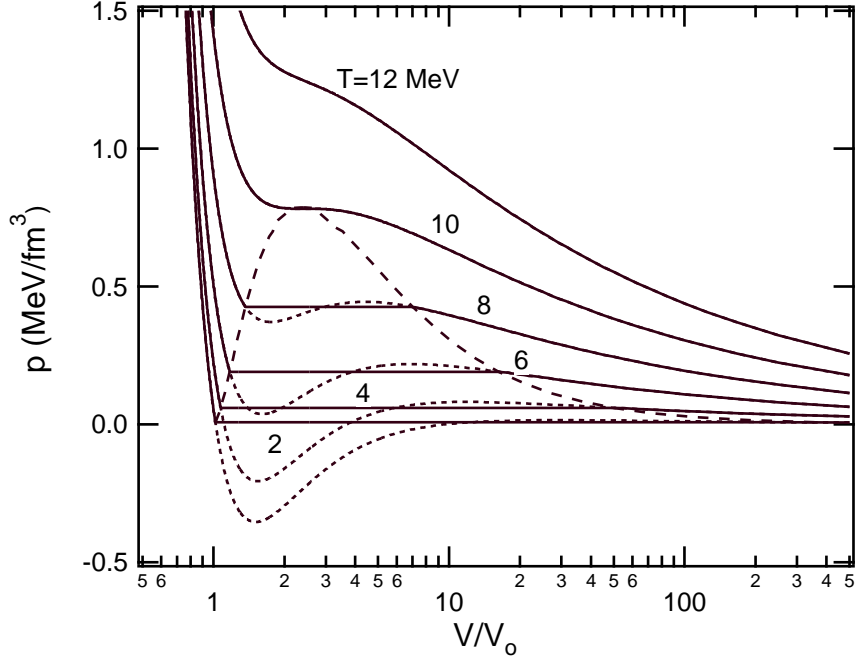}
\caption{\label{fig:isotherms} Isotherms calculated for an
$^{197}$Au-like system. Dotted lines illustrate isotherms for
hypothetical single-phase matter, while the dashed line visualizes
the boundary of the liquid-gas coexistence domain.}
\end{figure}

\begin{figure}
\includegraphics{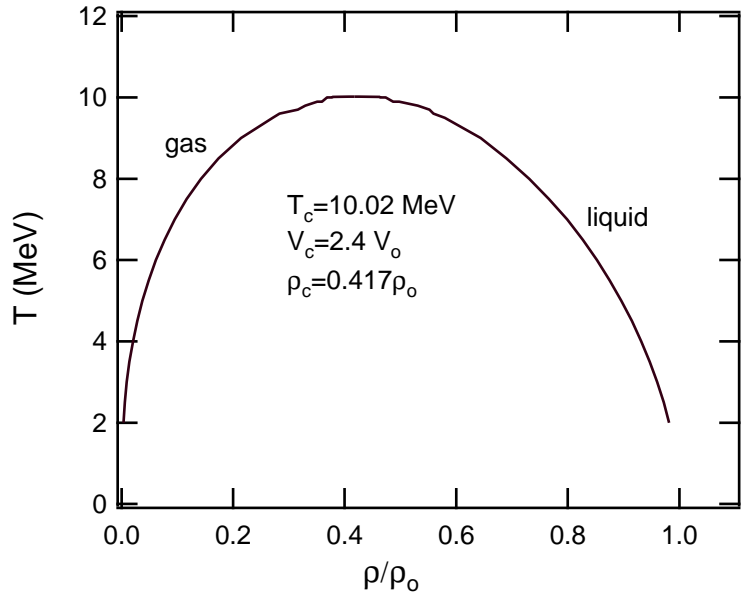}
\caption{\label{fig:coexist} Liquid-gas coexistence line in the
temperature {\it versus} matter density representation, as
predicted by the present formalism.}
\end{figure}

\begin{figure}
\includegraphics{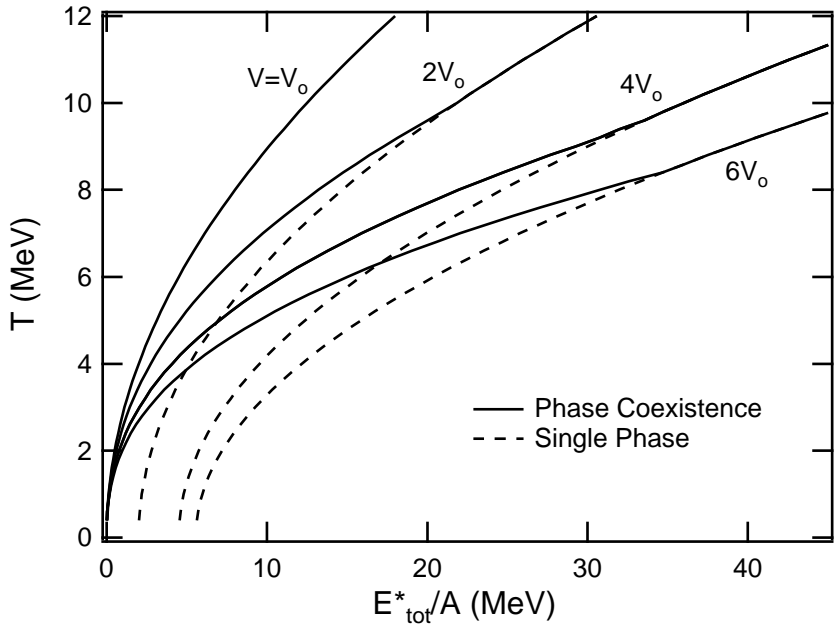}
\caption{\label{fig:caloric_curve} Isochoric caloric curves for
the boxed pseudo-Fermi matter, calculated for different volumes of
the confining box, as indicated by labels.}
\end{figure}

Results of a numerical minimization of the free energy of a
two-phase system are shown in
Figs.~\ref{fig:isotherms}-\ref{fig:critexp}. Fig.~1 presents
system isotherms, as predicted by the present formalism, in a
representation of system pressure $p$ versus system volume $V$.
Note that Eqs.~\ref{eq:free_energy_gas} and
\ref{eq:free_energy_liquid} represent single-phase states as
special cases. A pure liquid/gas state is thus among the possible
outcomes of the calculations with $A_{gas/liquid}=0$. As seen in
Fig.~\ref{fig:isotherms}, the isotherms feature segments
representing both, pure liquid and pure gas and, notably, the
liquid-gas coexistence ``plateaus''. The presence of coexistence
plateaus does not come as a surprise as the harmonic interaction
term of Eq.~\ref{eq:compress} has the salient characteristics of a
Van der Waals interaction.

It is worth noting that the coexistence plateaus in this
calculation result naturally from the actual minimization of the
free energy and are not obtained via the well-known
phenomenological Maxwell construct. Isotherms for hypothetical
single-phase states are shown as dotted lines. These latter
isotherms feature domains of spinodal instability characterized by
negative compressibility. The dashed line in
Fig.~\ref{fig:isotherms} illustrates the boundary of the
liquid-gas coexistence domain. The ``summit'' point of this curve
represents the critical point for the system and corresponds to a
critical temperature of $T_c\approx 10.0$ MeV. At temperatures
higher than $T_c$, the system can reside only in a single-phase,
vapor state.

A different representation of the liquid-gas coexistence line is
illustrated in Fig.~\ref{fig:coexist}. In this case, the
temperature $T$ is plotted versus the matter density, for points
along the boundary of the liquid-gas coexistence line (dashed line
in Fig.~\ref{fig:isotherms}). In the domain below this curve, the
system is in a two-phase state. In this coexistence domain, the
densities of gaseous and liquid phases at a given temperature are
depicted by the intersection points of the coexistence curve with
the horizontal line $T$=const. On the left shoulder of the curve
and in the domain further left to it, the system is in a pure
gaseous state, while on the right shoulder and in the domain
further right to it, the system is in a pure liquid state. The
difference between liquid and gaseous phases vanishes for
temperatures above $T=T_c=10.0$ MeV.

The search for the minimum free energy for a given temperature $T$
and given volume $V$ yields, along with the volumes
$V_{gas/liquid}$ and densities $\rho_{gas/liquid}$ of gaseous and
liquid phases, the total excitation energy $E^*_{total}$ of the
boxed matter. This allows one to construct the isochoric caloric
curves for the modelled system at different volumes of the
confining box. A set of such caloric curves is illustrated in
Fig.~\ref{fig:caloric_curve}. As expected, these curves do not
feature plateaus reported in various experiments,\cite{natowitz02}
but rather inconspicuous kinks at the locations on the boundary of
the coexistence domain.

\section{Critical Behavior}

One of the salient features of critical behavior in Van der Waals
systems is the presence of a singularity at the critical point.
Here, when the system temperature T approaches the critical
temperature $T_c$ from below, the difference between the densities
of coexisting liquid and gaseous phases vanishes according to a
power law
\begin{equation}
\label{eq:power_law} \rho_{liquid}-\rho_{gas}=C(T_c-T)^\beta,
\end{equation}

\begin{figure}
\includegraphics{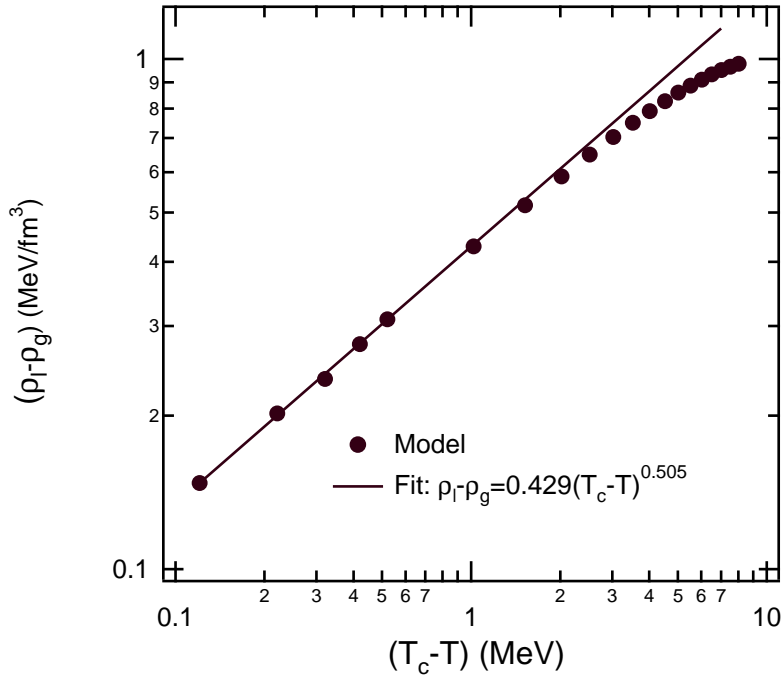}
\caption{\label{fig:critexp} Critical behavior of densities of
coexisting liquid and gaseous phases at temperatures close to
critical temperature $T_c$.}
\end{figure}

In Eq.~\ref{eq:power_law}, C is a constant and $\beta$ is the
critical exponent. The magnitude of the critical exponent can be
extracted conveniently by fitting a straight line to a
double-logarithmic plot of $(\rho_{liquid}-\rho_{gas})$ {\it
versus} $(T_c-T)$. Such a plot obtained in present calculations is
shown in Fig.~\ref{fig:critexp}. The plot features, indeed, an
approximately 2-MeV wide linear domain extending from the critical
temperature $T_c$ down to lower temperatures. A linear fit to this
domain allows one to extract the ``coordinates'' of the critical
point of $T_c$= 10.0 MeV, and $\rho_c/\rho_o$=0.42. Furthermore,
it yields a value of $\beta$=0.51 for the critical exponent.

Given the very schematic nature of the present formalism, the
predicted characteristics of the critical point are well within a
reasonable domain that can be inferred from more sophisticated,
but less transparent calculations.

\section{Discussion}

A simple formalism has been presented that allows one to model the
behavior of interacting pseudo-Fermi matter under the conditions
of controlled volume and temperature. The formalism is shown to
capture the essential physics of a first-order liquid-gas phase
transition. The calculation demonstrates the characteristics of
nuclear liquid-gas coexistence in a certain domain of system
parameters and the presence of a critical point. While such
characteristics are well expected, based on the the similarity of
the utilized form of interaction to that of the Van der Waals
interaction, the present formalism offers many didactical
benefits. For example, due to its simplicity it offers a clear
insight into physical phenomena it purports to describe and a
relatively strict test-bench for possible qualitative or
``hand-waving'' arguments. The results obtained, while almost
trivial, may alert one to possible challenges found by more
complete models. As an example, it is evident from Fig.~1 that
models of nuclear multifragmentation such as SMM \cite{smm} and
MMMC \cite{mmmc} tend to overpredict fragment multiplicities
significantly. This is so, because these models assume that the
liquid phase is always at ground-state density $\rho_o$. This
forces the pressure of the surrounding gas within the freezeout
volume to be higher than would be necessary to counter the lower
pressure of the expanded liquid at equilibrium density. A higher
gas pressure results in an overestimate of particle and fragment
multiplicities.

The formalism presented here leaves ample room for further
refinements such as a more strict modelling of the diluted Fermi
matter, or the inclusion of isotopic effects. At any rate, it
offers a convenient didactic tool to achieve a better
understanding of nuclear thermodynamics.

\begin{acknowledgments}
This work was supported by the U.S. Department of Energy grant No.
DE-FG02-88ER40414.
\end{acknowledgments}

\end{document}